\documentclass[seceq]{ptptex}
\notypesetlogo 
\title{On the Formation Age of the First Planetary System%        %You can use \\ for explicit line-break.
}

\author{%       %Use \scshape for the family name.
Tetsuya {\textsc Hara}\footnote{E-mail:hara@cc.kyoto-su.ac.jp}, 
Shuhei {\textsc Kunitomo}, Masanobu {\textsc Shigeyasu},
and Daigo {\textsc Kajiura}
}

\inst{%     %Affiliation, neglected when [addenda] or [errata].
Department of Physics, Kyoto Sangyo University, Kyoto 603-8555}

\abst{%         %This abstract is neglected when [addenda] or [errata].
Recently, it has been observed the extreme metal-poor stars in the Galactic halo, 
which must be formed just after Pop III objects. On the other hand, the first gas clouds 
of  mass $\sim 10^6 M_{\odot}$ are supposed to be formed at $ z \sim $ 10, 20, and 30 for the 
$1\sigma$, $2\sigma $ and $3\sigma$, where the density perturbations are assumed of  
the standard $\Lambda$CDM cosmology.  Usually it is approximated that the distribution 
of the density perturbation amplitudes is gaussian where $\sigma$ means the standard deviation.  
If we could apply this gaussian distribution to the extreme
 small probability, the gas clouds would be formed at $ z \sim $40, 60, and 80 for the 
$4\sigma$, $6\sigma$, and $8\sigma$ where the probabilities are approximately $3\times 10^{-5}$, $10^{-9},
$ and $10^{-15}$. Within our universe, there are almost $\sim 10^{16} \hspace{0.1cm} 
(\sim 10^{22}M_{\odot}/10^6M_{\odot})$ clouds
of mass $10^6M_{\odot}$.  Then the first gas clouds must be formed around $z \sim$ 80, where the time is 
$\sim$ 20 Myr $ \hspace{0.1cm}(\sim 13.7/(1+z)^{3/2}$ Gyr).  Even within our galaxy, there are 
$\sim 10^{5} \hspace{0.1cm} (\sim 10^{11}M_{\odot}/10^6M_{\odot})$ 
clouds, then the  first gas clouds within our galaxy must be formed around $z\sim 40$, where the time is 
$\sim$ 54 Myr $ \hspace{0.1cm}(\sim 13.7/(1+z)^{3/2}$Gyr).

   The evolution time for massive star ($\sim 10^2 M_\odot$) is $\sim 3$ Myr and the explosion of 
the massive supernova distributes the metal within a cloud.  The damping time of the supernova shock wave 
in the adiabatic and isothermal era is several Myr and stars of the second generation (Pop II) are formed 
within a free fall time $\sim 20$ Myr. Even if the gas cloud is metal poor, there is a lot of possibility 
to form the planets around such stars.  The first planetary systems could be formed within 
$\sim 6\times 10^7$ years 
after the Big Bang in the universe.  Even in our galaxies, the first planetary systems could be
 formed within $\sim 1.7\times 10^8$ years.  If the abundance of heavy elements 
such as Fe is small compared to the elements of C, N, O, the planets must be the one where
 the rock fraction is small.  It is interesting to wait the observations of planets around metal-poor stars.  
For the panspermia theory, the origin of life could be expected in such systems.

}

\begin{document}

\maketitle

\section{Introduction}
%Start your paper from here.

%\section*{Acknowledgements}
%We would like to thank ...........

  Recently it has been reported that the detection of the metal poor stars:
HE0107-5240 is a giant star ($36,000$ lyr, $ M \simeq 0.8M_\odot $),
 having just $1/200,000$ of the  solar metal abundance  \cite{Chi}\ :
HE1327-2326, discovered in 2005, is a star ($\geq 1,500$ lyr, $ M \simeq 0.7M_\odot $),
 having just 1/300,000 of the  solar metal abundance (the lowest known iron abundance to date). \cite{Fre}\
  It has been speculated that these stars are the second generation, 
born out of the gas clouds which were polluted by the primordial Population III stars.

  In the following we estimate the formation age of the first stars and consider the formation of 
the first planetary systems. If we consider the panspermia theory seriously, it could be speculated 
that first life began in such primordial planetary systems.

There has been established that rocks can be ejected from planetary surface by colliding asteroids and comets.  
The Chicxulub crater event 65 Myr ago provides evidence of the collisional ejection process. 
 The meteorites size is estimated about 10 km in diameter.

 However it is not certain that the micro-organisms within 
the size (  $ \ge$  1cm) of meteorites are still viable for several Myr, the Earth origin 
meteorites could be transferred to the nearby stellar systems.  If we take it is viable,
 we should consider the panspermia theories more seriously.

\section{Gaussian Distribution}

It is now understood that the first luminous stars are formed through the contraction of gas cloud due
 to cooling  
into the dark matter potential where the 
amplitudes of density perturbation become order one.  Without the density perturbation of dark matter, 
baryonic gas  could not contract due to background homogeneity. The formation of primordial 
gas clouds has been investigated (Matsuda, Sato and Takeda 1969,\cite {Mat}\  Palla, Salpeter, and Stahler 1983) \cite{Pal}\ and 
recently numerically elaborated (Bromm, Coppi and Larson 2002, \cite{Bro}\
Yoshida et al. 2003). \cite{Yos}

The first gas clouds 
of  mass $\sim 10^6 M_{\odot}$ are supposed to be formed at $ z \sim $ 10, 20, and 30 for the 
$1\sigma$, $2\sigma $ and $3\sigma$, where the density perturbations are assumed of  
the standard $\Lambda$CDM cosmology (Nishi and Susa 1999). \cite{Nis}\  Usually the distribution of the perturbation amplitudes is 
approximated as 
gaussian where $\sigma$ means the standard deviation.

If we could apply this gaussian distribution to the extreme
 small probability, the gas clouds would be formed at $ z \sim $40, 60, and 80 for the 
$4\sigma$, $6\sigma$, and $8\sigma$ where the probabilities are approximately $3\times 10^{-5}$, $10^{-9},
$ and $6 \times 10^{-16}$. Within our universe, there are almost $\sim 10^{16} \hspace{0.1cm} 
(\sim 10^{22}M_{\odot}/10^6M_{\odot})$ clouds
of mass $10^6M_{\odot}$.  Then the first gas clouds must be formed around $z \sim$ 80, where the time is 
$\sim$ 20 Myr $ \hspace{0.1cm}(\sim 13.7/(1+z)^{3/2}$ Gyr).  Even within our galaxy, there are 
$\sim 10^{5} \hspace{0.1cm} (\sim 10^{11}M_{\odot}/10^6M_{\odot})$ 
clouds, then the  first gas clouds within our galaxy must be formed around $z\sim 40$, where the time is 
$\sim$ 54 Myr $ \hspace{0.1cm}(\sim 13.7/(1+z)^{3/2}$Gyr). 
 The probability for each sigma is displayed in the table.

\vspace{0.4cm}
\begin{center}
%\catcode'?=\active \def?{phantom{00}}
 \begin{tabular}{l|r|r|r} \hline 
  \rule[0.2cm]{0.0cm}{0.2cm}   \hspace{0.5cm} $\sigma$ \hspace{0.3cm}  &   \hspace{1.0cm} probability \hspace{0.5cm}
 & \hspace{0.7cm}$\sigma$\hspace{0.4cm} &\hspace{1.0cm}  probability \hspace{0.7cm} \\  [4pt] \hline 
 \rule[0.2cm]{0.0cm}{0.2cm}  \hspace{0.4cm}  1   &   0.158   \hspace{1.1cm}   &   6 \hspace{0.3cm}  
& $ 9.90\times 10^{-10}$\hspace{0.5cm} \\  
   \hspace{0.5cm}  2   &  $ 2.27\times 10^{-2}\hspace{0.4cm} $  &   7 \hspace{0.3cm}  & $1.28\times 10^{-12}$\hspace{0.5cm} \\
   \hspace{0.5cm}  3   &  $ 1.35\times 10^{-3}\hspace{0.4cm} $  &   8 \hspace{0.3cm}  & $6.25\times 10^{-16}$\hspace{0.5cm} \\  
   \hspace{0.5cm}  4   &  $ 3.17\times 10^{-5} \hspace{0.4cm}$  &   9 \hspace{0.3cm}  & $1.13\times 10^{-19}$\hspace{0.5cm} \\
   \hspace{0.5cm}  5   &  $ 2.87\times 10^{-7} \hspace{0.4cm}$  &  10 \hspace{0.3cm}  & $7.66\times 10^{-24}$\hspace{0.5cm} \\ \hline
  \end{tabular}
\end{center} 
\vspace{0.6cm}
 Although the probability will increase if we take into account the non-gaussianity of the density perturbations
(Bartolo et al. 2004), \cite{Bar}\
we did not consider the effect.

\section{Formation of Population III stars}

It is now understood that the first luminous stars are formed through the following stages such as 
Jean's instability, free fall 
and Kelvin contraction.  Comparing the free fall time 
and cooling time (mainly $H_2$ cooling),  the gas clouds of 
$\sim 10^6M_{\odot}$ become Jean's instability and contracts into the dark matter potential as stated before.

In the following, we take $z \sim 80$ as the representative age for the first gas cloud formation, where
the time includes the contraction of the gas cloud (Susa, 2002).\cite{Sus}  
The age at $z \sim 80 $ is 
\begin{equation}
 t_{z}\sim \frac {13.7\rm{Gyr}}{(1+z)^{3/2}}\simeq 1.91\times 10^7 \left(\frac {z}{80} \right)^{-3/2}
\rm{yr}.
\end{equation}

During the contraction of gas cloud, 
the central part contracts almost free fall and becomes opaque (Omukai and Nishi 1998).\cite{Omu} 
The massive star of $\sim 10^2M_{\odot}$ 
is formed there.  In the opaque stage where the configuration becomes stable, radiation radiates 
the gravitational potential energy. The time scale of Kelvin contraction is  
\begin{equation}
 t_{K}\sim \frac {\rm{G} M^2}{2R}/L \simeq 2\times 10^7 \left(\frac {M}{100 M_{\odot}} \right)^{2}
\left(\frac {R}{10 R_{\odot}} \right)^{-1}
\left(\frac {L}{L_{\odot}} \right)^{-1}\rm{yr}.
\end{equation}
Although the formation age is a little bit different, the detailed contraction process of primordial 
gas clouds are numerically investigated and the results are not so much different
(Omukai and Nishi 1998,\cite{Omu}\   Bromm, Coppi and Larson 2002, \cite{Bro}\
Yoshida et al. 2003).\cite{Yos}\  

   The evolution time for massive star ($\sim 10^2 M_\odot$) is $\sim 3$ Myr and the explosion of 
the massive supernova distributes the metal within a 
cloud.

\section{Formation of Planetary System}

The damping time of the supernova shock wave in the adiabatic and isothermal era 
 is several Myr.  Here we assume it takes the time scale of the following free fall time.

The contraction and the free fall time of the cloud around $z \sim 80 $ is
\begin{equation}
 t_{ff}\sim \frac {1}{\sqrt{\pi {\rm G} \rho}}\simeq 2\times 10^7 \left(\frac {z}{80} \right)^{-3/2}
\rm{yr}.
\end{equation}
(the gas cloud is assumed to collapse at $z \sim 80 $ so the mean density at turn around is two times of 
the background density at $z \sim 1.6 \times  80 $ (Susa, 2002)).\cite{Sus}\
If the mass of gas cloud is smaller than $\sim 10^6 M_{\odot}$, the explosion energy of SN is 
greater than the gravitational binding energy. 
So the mass of the gas cloud must be greater than  $\sim 10^6 M_{\odot}$.  
Then the increase of metal abundance $\delta Z$ in the cloud is order of 
$\delta Z\sim 10^{-5} \sim 10M_{\odot}/{10^6 M_{\odot}}$.

After the contraction of metal contaminated gas cloud, the stars of the second 
generation (Pop II) are formed within a free fall time $\sim 20$ Myr.  Then the first planetary systems 
could be supposed to be formed within $\sim 6\times 10^7$ years after the Big Bang. 

The same arguments could be applied to the case of $z\sim 40$ for 4$\sigma$ where the collapsed 
time is $\sim$ 54 Myr. After the explosion of the gas cloud, the expected elapsed time for the adiabatic, 
isothermal expansion 
and the free fall time is 
$t_{ff}\sim 6\times 10^7 \left( z/40 \right)^{-3/2} \rm{yr}. $  
 So the first planetary systems for this case
could be supposed to be formed within $\sim 1.7\times 10^8$ years after the Big Bang.

 The Population II stars reside in Halos and globular clusters. Even stars in globular clusters could take 
planetary systems for enough time (several Gyr),  it is expected the life could start there.

 The fragmentation of the supernova shocked shell is investigated by Salvaterra, Ferrara and Schneider (2004)
 \cite{Sal}\  and they estimated the instability of the shell sets in $ 0.2 \sim  50 $ Myr after the explosion, 
depending on the explosion energy 
and the density of the surrounding medium.  If we adopt their estimation of the short time scale  $ \sim 0.2 $ Myr, 
the formation age of the first planetary system will become as fast as $ \sim 43 $ Myr after the Big Bang. 
They also discussed the metallicity of the fragments 
through the mixing of the heavy 
elements into the swept up matter in the shell. Although there are many uncertainty mechanism,
 they estimated the mean metallicity of the fragments 
$ 10^{-3.5} Z_{\odot} \leq  Z \leq 10^{-2.6} Z_{\odot} $. If we include their estimation, 
the metallicity 
of the secondary stars will much increase and the formation of the planetary systems around 
the secondary stars will be much expected in the earlier stage of the cosmography.

\section{Conclusions and Discussion}

The formation of the first planetary system is related to gaussian distribution of dark matter perturbation.
Within our universe, there are almost $\sim 10^{16} \hspace{0.1cm} 
(\sim 10^{22}M_{\odot}/10^6M_{\odot})$ clouds
of mass $10^6M_{\odot}$.  Then the first gas clouds must be formed around $z \sim$ 80, where the time is 
$\sim$ 20 Myr $ \hspace{0.1cm}(\sim 13.7/(1+z)^{3/2}$ Gyr). Considering the evolution, SN explosion, expansion,
 and contraction, the first planetary systems with metals 
could be supposed to be formed within $\sim 6\times 10^7$ years after the Big Bang.

  Even within our galaxy, there are 
$\sim 10^{5} \hspace{0.1cm} (\sim 10^{11}M_{\odot}/10^6M_{\odot})$ 
clouds. As stated before the  first gas clouds within our galaxy must be formed around $z\sim 40$, 
where the time is $\sim$ 54 Myr $ \hspace{0.1cm}(\sim 13.7/(1+z)^{3/2}$Gyr). 
The first planetary systems for this case
could be supposed to be formed within $\sim 1.7\times 10^8$ years after the Big Bang. 

It is interesting to wait the observations of planets around metal-poor stars in 
the halos and globular clusters.
For the panspermia theory, the origin of life could be expected in such systems.  The ejected rocks from 
such planets could transfer micro-organism within galaxies (Wallis and Wickramasinghe 2004). \cite{Wal}\
   Even in our solar system, there must be many types of life in Mars, Europa, Ganymede, Io, Uranus,  
Neptune, Pluto, and Titan.  It is interesting to wait the investigation of these planets and 
satellites.

%\appendix
%\section{First Appendix} %Empty argument \section{} yields `Appendix'. 
%
%\section{Second Appendix}

\end{document}